\documentclass[10pt]{article}
\usepackage[cp1251]{inputenc}
\usepackage[english]{babel}
\usepackage{graphicx}
\usepackage{amssymb}
\usepackage{amsmath}
\usepackage{amsthm}
\usepackage{amsfonts}
\usepackage{amssymb}
\title{{\bf \Large  Fractional Einstein-Hilbert Action Cosmology}\\
{\normalsize ~~{\bf V.\,K. Shchigolev}\thanks{E-mail:
vkshch@yahoo.com}}\\
{\small {\it Ulyanovsk State University, 42 L. Tolstoy Str.,
Ulyanovsk 432000, Russia}}\\
\vspace{2mm}
\small \begin{quote}{\bf Abstract} --  We propose a new type of cosmological model derived from the fractional variational principle when it is applied to the gravitational sector of action functional. In contrast to the fractional cosmological model developed earlier by the author from a  fractional total action, in our new model the continuity equation remains valid in its usual form.  For this model, a lot of exact solutions  are obtained from a specific {\it ansatz}  which is proposed for the cosmological term in this paper. Several examples arising from the given variations of the Hubble parameter with time  are provided. Besides, we suggest an original interpretation of the main equations for our model. It supposes that the effective cosmological term could arise as a result of kinematical induction  through the non-zero Hubble parameter. With the help of particular example,  we demonstrate how this approach could lead our model quite closer to the real behavior of  the universe.
 \\
\vspace{2,5mm}
{\bf PACS numbers}: 98.80.-k; 98.80.Jk; 04.20.Jb.\\
{\bf Key words}: Cosmological Models, Fractional Einstein-Hilbert Action, Exact Solutions, Accelerated Expansion, Induced Cosmological Term.\\
\end{quote}}
\date{}

\begin{document}
\maketitle
\vspace{-20mm}
\section{Introduction}

A number of remarkable features of the fractional action cosmology and the theory of fractal universe have inspired a certain interest  to such a sort of theories over the last few years (see for example \cite{1Shchigolev}  - \cite{Yerokhin} and references therein). All studies of implementation of the fractional measure in cosmology recently published, with all their differences in motivation and with some differences in the basic equations, have a number of common features. First of all, both of these theories are derived from the similar integrals of action  built on the measure $d \varrho(x)$ which is the  Lebesgue-Stieltjes measure generalizing the standard 4-dimensional measure $d^4x$. This is not so surprising  fact because, as noted in \cite{1Calcagni}, integrals on net fractals can be approximated by the left-sided Riemann-Liouville fractional integral. The second common feature of these two cosmological theories is that their continuity equations almost coincide and include similar terms,  perturbing the energy conservation law. A perturbed continuity equation is not some specific property of the fractional (fractal) cosmology but is almost common feature of many modifications of the gravity theory.

The desire to preserve the continuity equation in the framework of the fractional action cosmology in the same form, in which it is written in General Relativity, leads to finding out the reasons of it's perturbations. In the case of fractal cosmology, the formal reason is the  Lebesgue-Stieltjes measure in the action of matter. Hence, the possibility to save the continuity equation in its canonical form is real, if one  assumes that the total action in the condition of minimal coupling can be equipped with two different measures. In this case, the action for matter should be equipped with  the standard measure, but in the gravitational sector one has to maintain the Stieltjes measure given the possible fractal properties of space-time.
The idea of two different measures in itself is not new (see for example \cite{Guendelman} and references therein). This idea was motivated by some other concepts, so it was realized by other way,  and it never has been used in the context of fractional cosmology.

Our key idea is to keep the usual form for the continuity equation  in the framework of fractional variational principle. We are going to achieve this aim by means of retaining the concept of fractional order for the action functional only with respect to space-time.
In other words, the purpose of this work is to build a new cosmological model which follows from the  effective action of a fractional order for gravity. Making use of  this assumption, we derive the basic equations of our model.  The problem of finding  solutions for  the modified equations in  fractional action cosmology  is no less complicated problem than the same one in the standard cosmology. We offer different methods of solving for these equations in order to obtain the exact solutions and to discuses  their properties. Unfortunately, hitherto we know relatively small number of exact solutions for the fractional action  (and as well in fractal) cosmology . Most of these solutions are obtained from  the specified regimes of evolution of scale factor \cite{Debnath} - \cite{Jamil},  or by means of some ansatz  as been suggested in \cite{2Shchigolev}.  In this paper, solutions for the new fractional cosmological model are obtained  from the special ansatz for cosmological term, and with the help of other assumptions.

Moreover, we propose to arrange the set of main equations in such a way  that the effective  $\Lambda$ - term could be treated as a kinematically induced (by the Hubble parameter) cosmological term.   We show in a specific example that our model based on this proposal  could lead to some rather realistic regimes of  expansion of the universe.

\section{The model equations}

In the approach developed by the author in \cite{1Shchigolev} , \cite{2Shchigolev}, the action integral $S_L [q]$ for the Lagrangian density $L(\tau, q(\tau), \dot q(\tau))$ is written as a fractional Riemann-Stieltjes integral \cite{Uchaikin}:
\begin{equation}
\label{1} S_L [q_i]=\frac {1}{\Gamma
(\alpha)}\int\limits_{t_0}^{t} L(\tau,q_i (\tau),\dot
q_i(\tau))(t-\tau)^{\alpha-1} d\tau ,
\end{equation}
with the integrating function ${\displaystyle g_t(\tau) = \frac{1}{\Gamma (1 + \alpha)} [t^{\alpha} - (t-\tau)^{\alpha}]}$. The latter has the following scaling property: $g_{\mu t}(\mu \tau) = \mu^{\alpha} g_t(\tau), ~ ~ \mu > 0$.

Here, we consider a new cosmological model which can be derived from a variational principle for the Einstein-Hilbert action $\displaystyle S_{EH}=\frac{M_{P}^2}{2}\int \sqrt{-g}\,d^{\,4}x (R-2\Lambda)$, where $M_P^{-2}=8\pi G$ is reduced Planck mass. Following the definition (\ref{1}), the modified fractional effective Einstein-Hilbert action in a spatially flat Friedmann-Robertson-Walker metric, $$\displaystyle ds^2 = N(t)^2 dt^2 - a^2(t)\delta_ {ik} dx^i dx^k ,$$
where $N$ is the lapse function and $a(t)$ is a scale factor, is represented by the fractional integral as follows \cite{2Shchigolev}
\begin{equation}
\label{2} S_{EH}^{\alpha}=\frac {1}{\Gamma
(\alpha)}\int\limits_{0}^{t} \frac{3}{8\pi G}
\left(\frac{a^2\ddot a}{N}+\frac{a\dot a^2}{N}-\frac{a^2\dot a
\dot N}{N^2}-Na^3\frac{\Lambda}{3}\right)(t-\tau)^{\alpha-1} d\tau~.
\end{equation}
Assuming the matter content of the universe is minimally coupled to gravity, the total action of the system is
$$S_{total}^{\alpha}=S_{EH}^{\alpha}+S_{m},$$
where the effective action for matter can be represented by the usual expression, which follows from the matter action  with the standard measure $S_{m}=\int {\cal L}\sqrt{-g} d^4x$:
\begin{equation}
\label{3}
S_{m}=\int {\cal L}_m N a^3 d\,t.
\end{equation}
Making use of  the fractional variational procedure, developed in \cite{1Shchigolev} and  applied to $S_{total}$, and taking into account Eqs. (\ref{2}) and (\ref{3}), we can derive the following dynamical equations for our model:
\begin{eqnarray}
3 H^2+3\frac{(1-\alpha)}{t}H=t^{1-\alpha} \rho+\Lambda{~,}\label{4}\\
2\dot H + 3 H^2+2\frac{(1-\alpha)}{t}H +
\frac{(1-\alpha)(2-\alpha)}{t^2}=-t^{1-\alpha} p+\Lambda{~,}\label{5}
\end{eqnarray}
where $H(t)=\dot a/a$ is the Hubble parameter, and  we put $8\pi G\, \Gamma
(\alpha)=1$ for simplicity. A remarkable feature of this model is that the continuity equation is written in the usual form,
\begin{equation}
\dot \rho+3H(\rho+p)=0,\label{6}
\end{equation}
expressing the standard energy conservation law for a perfect fluid, just the same
as in the cosmological theory of General Relativity. It can be shown that Eqs. (\ref{4}) and (\ref{5}) imply the continuity equation (\ref{6}) in the case $\alpha \ne 1$, if the following equation is valid:
\begin{equation}\label{7}
\dot H - 2\frac{(2-\alpha)}{t}H
= \frac{t^{\displaystyle 2-\alpha}}{3(1-\alpha)}\,\frac{d }{d t}\Big(t^{\displaystyle \alpha - 1}\Lambda\Big),
\end{equation}
Let us note that the latter equation is written after dividing by non-zero multiplier $(1-\alpha)$. It is why the constancy of cosmological term, $\Lambda = constant$, in the limit $\alpha = 1$ is followed from this equation.
Using Eqs. (\ref{4}) and (\ref{5}), we can obtain the equation of state (EoS) of matter as follows:
\begin{equation}\label{8}
w_m = \frac{p}{\rho}= -1-\frac{2}{3}\,\frac{\displaystyle \frac{\dot H}{H^2}-\frac{1-\alpha}{2(tH)}+\frac{(1-\alpha)(2-\alpha)}{2(tH)^2}}{\displaystyle 1+\frac{1-\alpha}{(tH)}-\frac{\Lambda}{3 H^2}}.
\end{equation}
It should be noted also that the effective EoS is a dynamical characteristic of the model and  gains a new definition represented by the formula (\ref{8}). The deceleration parameter,
\begin{equation}\label{9}
q = -\frac{a^2\, \ddot a}{\dot a^2} = -1-\frac{\dot H}{H^2},
\end{equation}
is defined just as in the standard cosmology, being a kinematical parameter  of the model \cite{Starobinsky}.

Let us now obtain a class of exact solutions to the equation (\ref{7}) in the case $w_m \ne -1$, assuming that  the cosmological term $\Lambda(t)$  is given by some ansatz.  First of all, we make the following substitution into Eq. (\ref{7}):
\begin{equation}\label{10}
x=\ln (t/t_0) \Leftrightarrow t=t_0 \exp(x);\,\,\, Y(t) = t\, H(t),
\end{equation}
where $t_0> 0$ is a constant. As a result, this equation can be rewritten as
\begin{equation}\label{11}
Y' -(5-2\alpha) Y = \frac{t_0^2}{3(1-\alpha)} e^{\displaystyle(3-\alpha)x}\Big( e^{\displaystyle-(1-\alpha)x}\Lambda\Big)',
\end{equation}
where the prime denotes the derivative with respect to $x$. Obviously, a lot of exact solutions can be obtained for our model when its  $\Lambda$ - term satisfies the following equation:
\begin{equation}\label{12}
\Big[ e^{\displaystyle-(1-\alpha)x}\Lambda(x)\Big]' = \frac{3(1-\alpha)}{t_0^2} e^{\displaystyle(\alpha-3)x}\Big[c_1 Y'(x) + c_2 Y(x)+c_3+F(x)\Big],
\end{equation}
 where $c_i$ are constants, and $F(x)$ is an arbitrary smooth function. Substituting (12) into Eq. (11), we obtain  Eq. (\ref{12}) in the form
\begin{equation}\label{13}
K Y'(x) - L Y(x) = M + F(x),
\end{equation}
where the coefficients equal
\begin{equation}\label{14}
K = 1-c_1,\,\,\,L = 5-2\alpha +c_2,\,\,\,M = c_3.
\end{equation}
It is not difficult to find a generic solution for Eq. (\ref{13}):
\begin{equation}
Y(x) = \displaystyle-\frac{M}{L} + \Big[\frac{1}{K}\int F(x)e^{\displaystyle-Lx/K}d\,x +const\Big]e^{\displaystyle Lx/K},\nonumber
\end{equation}
that allows us to write the following expression for the Hubble parameter:
\begin{equation}\label{15}
H(t) = \frac{1}{t} \cdot \left[\frac{H_0 L}{K} \cdot t^{\displaystyle L/K}  -\frac{M}{L} +\frac{t^{\displaystyle L/K}}{K}\int F(t)t^{\displaystyle -L/K-1}d\,t\right],
\end{equation}
where we have introduced a new integration constant $H_0$. The generic character of generating function $F(t)$ provides an ample opportunity to study the evolution of the model. Nevertheless,  here we will focus on the simplest  case $F(t)\equiv0$. In this case,   the law of evolution of the scale factor follows from (\ref{15}) in the form:
\begin{equation}\label{16}
a(t) = a_0 t^{\displaystyle -M/L} \exp \left\{H_0 t^{\displaystyle L/K}\right\}
\end{equation}

As examples, we consider three cases for the constants  $c_i$   in Eq. (\ref{12}), writing it in the original variables according to (\ref{10}):
\begin{equation}\label{17}
\frac{d}{d t}\Big[ t^{\alpha-1}\Lambda(t)\Big] = 3(1-\alpha)t^{\alpha-4}\Big[c_1 t^2 \dot H(t) + (c_1+ c_2) t H(t)+c_3\Big].
\end{equation}

(1) Let all $c_i = 0$. From Eq. (\ref{17}), then it follows that $\Lambda = \Lambda_0\, t^{\displaystyle 1-\alpha}$, where $\Lambda_0 > 0$ is a constant of integration. Due to Eq. (\ref{14}), we have for the coefficients in the solution (\ref{15}) and (\ref{16}): $K=1,\, L=5-2\alpha,\, M=0$.

(2) Let us now suppose that $c_1 = c_2 = 0$ and $c_3 = -\beta (3-\alpha)/3(1-\alpha) < 0$, where $\beta > 0$ is a constant parameter. Integrating Eq. (\ref{17}) and assuming that a constant of integration equals zero,  we obtain one of the well known expression for the cosmological term: $\Lambda = \beta/t^2$.

(3) Suppose that in this case $c_1 = \beta/3(1-\alpha),\, c_2=-c_1(3-\alpha)$ and $c_3 = 0$, where $\beta > 0$ is a positive constant. Integrating Eq. (\ref{17}), and again assuming that a constant of integration is zero, we obtain other well-known expression for the cosmological term: $\Lambda = (\beta/t)H(t)$.

\subsection{Models with a constant EoS of matter}

One can verify that   the EoS of  matter (\ref{8}) depends on time in all cases considered above. To study the behavior of our model in the case of a constant EoS, i.e.  $w_m = constant$, it is convenient to rewrite equation (\ref{8}) with respect to the Hubble parameter in the following form:
\begin{equation}\label{18}
2\dot H + 3(1+w_m) H^2  + (2+3 w_m) \frac{(1-\alpha)}{t}H
+\frac{(1-\alpha)(2-\alpha)}{t^2}
= (1+w_m)\Lambda.
\end{equation}
The well-known solution for Eq. (\ref{6}) can be represented in the standard form $\rho =
\rho_0 a^{\displaystyle -3(1+w_m)}$. Note that Eq. (7) is still valid, allowing to derive of Eq. (6) from Eqs. (\ref{4}) and (\ref{5}). The substitution of $\Lambda$  from Eq. (18) into Eq. (7) leads to the following equation for the Hubble parameter:
\begin{eqnarray}\label{19}
2t\ddot H +6(1+w_m)tH\dot H-3(1-\alpha)\dot H + (1-\alpha)(2-\alpha)(4+3w_m)\frac{H}{t}- \nonumber\\-
3(1+w_m)(1-\alpha)H^2- (1-\alpha)(2-\alpha)(3-\alpha)\frac{1}{t^2}=0.
\end{eqnarray}
Exact analytical solutions to this equation can be obtained only for some particular values of $w_m$ and $\alpha$,  or one can obtain these solutions numerically. It should be kept in mind here that the order of equation is increased as a result of substitution of Eq.(18) into Eq. (7), and some solution may not satisfy the original equation. Therefore, the solution for Eq. (\ref{19}) should be inserted into Eqs. (\ref{4}), (\ref{5}) together with $\rho = \rho_0 a^{\displaystyle -3(1+w_m)}$. Obviously, this problem is of a certain interest, and we are going to study it later.

One can see that  Eq. (\ref{18}) for the quasi-vacuum matter $w_m = -1$  reduces to the following equation
\begin{equation}\label{20}
\dot H - \frac{1-\alpha}{2 t}H +\frac{(1-\alpha)(2-\alpha)}{2t^2}= 0\,,
\end{equation}
from which it follows that the Hubble parameter varies with time as
\begin{equation}
\label{21} H = \frac{C_{\alpha}}{t}+ H_0\,
t^{\displaystyle\frac{1-\alpha}{2}}~.
\end{equation}
Here, and henceforth in this paper,
$C_{\alpha}=\displaystyle\frac{(1-\alpha)(2-\alpha)}{(3-\alpha)}\,$, and
$H_0$ is a positive constant of integration. The latter equation yields the scale factor as
\begin{equation}
\label{22} a = a_0\,\, t^{\displaystyle C_{\alpha}}\exp
\left(\frac{3-\alpha}{2}H_0 t^{\displaystyle \frac{3-\alpha}{2}}
\right) ~.
\end{equation}
As an illustration, the graphs of functions $H(t)$ and $a(t)$ for $\alpha = 0.8$ are shown in Fig. 1. The usual result, $\rho=\rho_0=constant$, follows from Eq. (\ref{6}) with $w_m=-1$.
Therefore, the cosmological term as a function of time can be found from Eq. (\ref{4}) in the following form
\begin{equation}
\label{23} \Lambda(t) = (3 H_0^2-\rho_0) t^{\displaystyle 1-\alpha}+
3H_0\,C_{\alpha}\frac{(7-3\alpha)}{(2-\alpha)}\phantom{.}t^{\displaystyle-\frac{(1+\alpha)}{2}}+
3C_{\alpha}^{2}\frac{(5-2\alpha)}{(2-\alpha)}\,t^{-2}.
\end{equation}
If $\rho_0 = 3 H_0^2$, the $\Lambda$ - term decreases with time, which is consistent with the observational data. Note that the same result the author obtained from the quasi-vacuum EoS of matter in the framework of cosmology with the  fractional total action  \cite{2Shchigolev}.
\begin{figure}[t]
\centering
\includegraphics[width=0.47\textwidth]{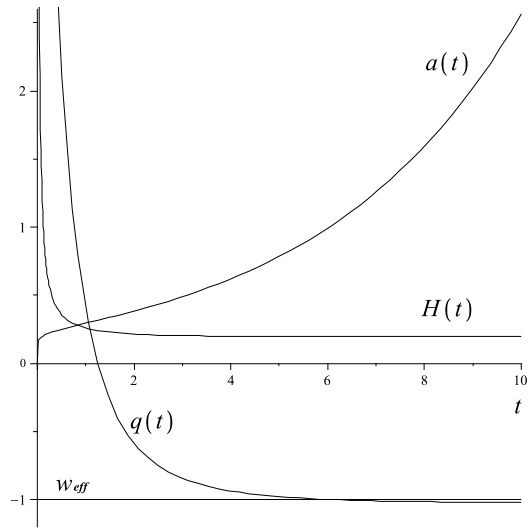}\\
\caption{The evolution of the scale factor $a$, the Hubble parameter $H$, the effective EoS $w_{eff}$ and the deceleration parameter $q$ in the model $w_m = -1$. Here, $\alpha=0.8$, and $H_0=0.15$.}
\label{fig1}
\end{figure}

\section{Models with a kinematically induced cosmological term}

Equation (\ref{7}) can be regarded as the equation for finding the cosmological term $\Lambda(t)$.  It can be solved in quadratures:
\begin{equation}\label{24}
\Lambda(t)=3(1-\alpha)\left[\frac{H(t)}{t}-(2-\alpha)t^{1-\alpha}\int t^{\alpha-3} H(t)d\,t\right]+ \Lambda_0 t^{1-\alpha},
\end{equation}
where $\Lambda_0$  is a constant of integration. Substituting Eq. (\ref{24}) into the basic equations of our model (\ref{4}), (\ref{5}), we obtain the following set of equations:
\begin{equation}\label{25}
3H^2=t^{1-\alpha}\rho_{eff},
\end{equation}
\begin{equation}\label{26}
2\dot H+3H^2-\frac{1-\alpha}{t}H+\frac{(1-\alpha)(2-\alpha)}{t^2}=-t^{1-\alpha}p_{eff},
\end{equation}
where the effective energy density and pressure are represented by
\begin{equation}
\rho_{eff}=\rho+\Lambda_{eff},\,\,\,\,\,\,p_{eff}=p-\Lambda_{eff},\nonumber
\end{equation}
\begin{equation}
\Lambda_{eff}=\Lambda_0+\Lambda_{ind}=\Lambda_0-3(1-\alpha)(2-\alpha)\int t^{\displaystyle \, \alpha-3}H(t)d\,t.\label{27}
\end{equation}
The latter means that the effective cosmological constant is the sum of the cosmological constant $\Lambda_0$ and induced (by the Hubble parameter) cosmological term $\Lambda_{ind}$. Combining  Eqs. (\ref{25}), (\ref{26}), we can derive the continuity equation  for the effective values of the energy density and pressure as follows
$$
\dot \rho_{eff}+3H(\rho_{eff}+p_{eff})= -3(1-\alpha)(2-\alpha)t^{\displaystyle \alpha-3}H,
$$
which reduces to the continuity equation for matter (\ref{6}) due to Eq. (\ref{27}). Thus, the set of equations (\ref{25}), (\ref{26}) consists of two independent equations, and can determine the dynamics of our model.

It can be assumed that the effective parameters  $\rho_{eff}$ and $p_{eff}$  satisfy
some effective EoS. For simplicity, in this study we consider the effective barotropic fluid, for which the effective EoS  follows from Eqs. (\ref{25}) and (\ref{26}) in the form:
\begin{equation}\label{28}
w_{eff} = \frac{p_{eff}}{\rho_{eff}} = -1 -\frac{2}{3}  \frac{\dot H}{H^2}+\frac{1-\alpha}{3(tH)}\left[1-\frac{2-\alpha}{(tH)}\,\right].
\end{equation}
Then from Eqs. (\ref{25}) and (\ref{26}), we obtain that
\begin{equation}
\rho = 3 t^{\displaystyle \alpha-1} H^2 - \Lambda_{eff},\,\,\,
p = 3 t^{\displaystyle \alpha-1} H^2 w_{eff} + \Lambda_{eff}.\label{29}
\end{equation}
Assuming the matter obeys also a barotropic EoS $p= w_m \rho$, we obtain the following equations relating the barotropic indexes of the effective fluid and matter:
\begin{equation}\label{30}
w_m=\frac{p}{\rho}=-1 +\frac{\displaystyle 1+w_{eff}}{\displaystyle 1-\frac{\Lambda_{eff}}{3H^2}t^{\displaystyle 1-\alpha}},\,\,\,\,\, w_{eff}=w_m -(1+w_m)\frac{\Lambda_{eff}}{3H^2}t^{\displaystyle 1-\alpha}.
\end{equation}
Hence it is easy to derive an interesting property of the quasi-vacuum state: only for this state, the equality $w_{eff}=w_m$ is valid, that is $w_{eff} =-1 \Leftrightarrow w_m = -1$. The latter means that if the model assumes crossing the phantom divide  $-1$, then $w_{eff}$ and $w_m$ can cross it only simultaneously.

In addition,it is easy to verify that  Eq. (\ref{29}) for the energy density and pressure identically satisfies the continuity equation (\ref{6}) due to the definition (\ref{27}). Therefore, if we assume that some field is the source of gravity, then its field equation is already satisfied in the form of  Eq. (\ref{6}). So for the reconstruction of the field and its potential,   only Eq. (\ref{29}) with the appropriate energy density and pressure should be solved. Of course, there are many other formulations of the problem.

\subsection{Models with a given Hubble parameter $H(t)$}

As known, the set of cosmological equations could be solved with the help of   phenomenologically given  law for $H(t)$. To illustrate this approach in our case, we assume some cosmological scenarios which are studied in the literature \cite{Debnath}.

(1) First of all, we consider the model in an inflationary scenario with  $H(t)=H_0$ and $a(t)=a_0 \exp(H_0 t)$, putting $M=0,\,\, L/K=1$ and $F(t)=0$ in Eq. (\ref{15}) for the Hubble parameter, and in Eq. (\ref{16}) for the scale factor. Keeping in mind Eq. (14), we obtain the following relations for the set of coefficients $c_i$: $c_3 =0$, $c_1+c_2=2(\alpha-2)$. The substitution of $H(t)=H_0$ into Eq. (\ref{27}) yields  the effective cosmological term as
\begin{equation}\label{31}
\Lambda_{eff}=\Lambda_0+\frac{3(1-\alpha)H_0}{t^{\displaystyle \,2-\alpha}},
\end{equation}
and Eq. (\ref{28}) leads to the following effective EoS:
\begin{equation}\label{32}
w_{eff} = -1 + \frac{1-\alpha}{3H_0 t}\Big[1-\frac{2-\alpha}{H_0 t}\Big].
\end{equation}
It can be seen that the model crosses the phantom divide-line at $H_0 t=2-\alpha$ and asymptotically tends to it  as $H_0 t \to \infty$. According to Eqs. (\ref{30})-(\ref{32}), the EoS of matter varies with time as follows:
\begin{equation}\label{33}
w_{m} = -1 + \frac{(1-\alpha)}{3H_0 t}\frac{\displaystyle \Big(1-\frac{2-\alpha}{H_0 t}\Big)}{\displaystyle \Big(1-\frac{1-\alpha}{H_0 t}-\frac{\Lambda_0}{3 H_0^2} t^{1-\alpha}\Big)}.
\end{equation}
In Fig. 2, we show the evolution of  $w_m$, $w_{eff}$ and the deceleration parameter $q$ in this model. One can observe the transition of matter from a state with $w_m>0$ to the state of quintessence and the subsequent crossing of the phantom divide.

\begin{figure}[t]
\includegraphics[width=0.47\textwidth]{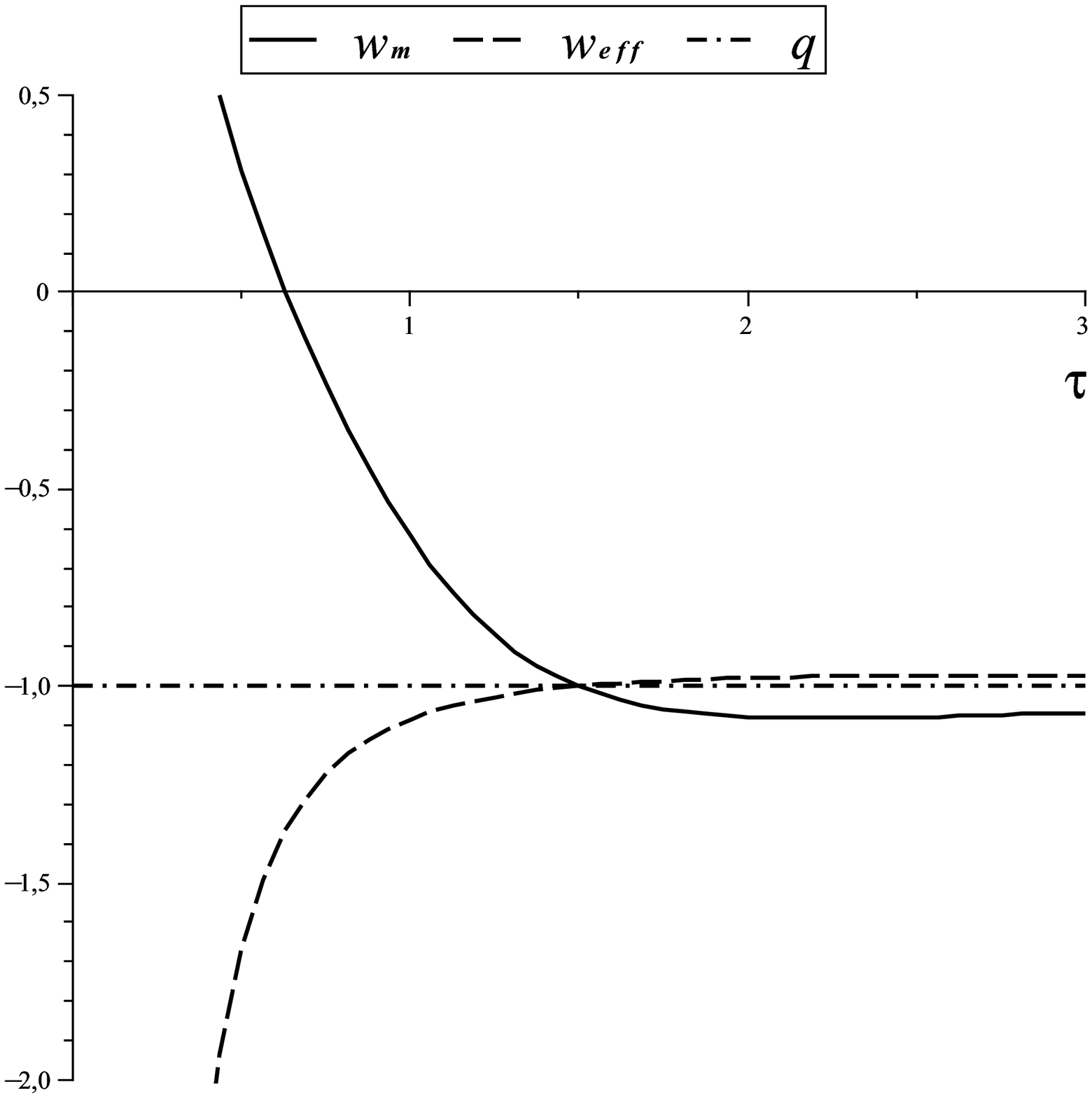} \hfill
\includegraphics[width=0.47\textwidth]{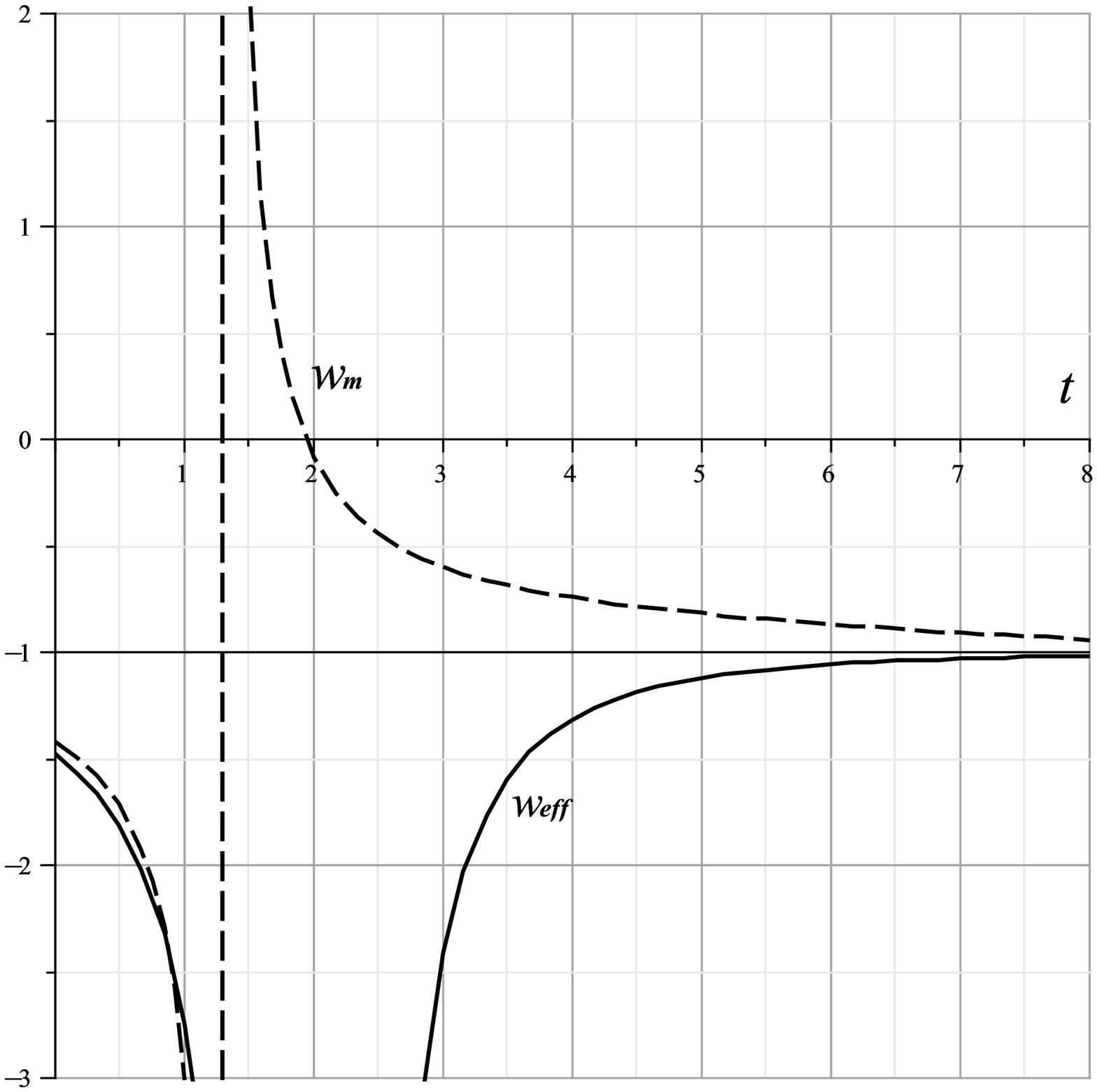}
\parbox[t]{0.47\textwidth}{\caption{The EoS of matter $w_m$, the effective EoS $w_{eff}$ and
the deceleration parameter $q$ in the model $H=H_0$ as the functions of dimensionless time $\tau = H_0 t$. Here,  $\alpha = 0.5,\,\,\,\Lambda_0/3H_0^{3-\alpha}=0.72$.}\label{fig2}} \hfill
\parbox[t]{0.47\textwidth}{\caption{The EoS of matter $w_m$ and the effective EoS $w_{eff}$ in the model of bouncing universe with $\epsilon = +1$. Here, $\alpha = 0.5,\,n=2,\, m=1,\, H_0=1, \,\Lambda_0=0.72$.}\label{fig3}}
\end{figure}

(2) We now consider the power law, assuming $H(t)=\displaystyle \frac{n}{t}$, where $n$ is a
positive constant. Then from (\ref{27}), we have
\begin{equation}\label{34}
\Lambda_{eff}=\Lambda_0+\frac{3C_{\alpha}n}{\displaystyle t^{3-\alpha}}.
\end{equation}
Note that in this case we arrive at a constant effective EoS. Indeed, substituting both $H(t)$ and the cosmological term (\ref{34}) into equation (\ref{14}), we get:
\begin{equation}\label{35}
w_{eff} = -1+\frac{3-\alpha}{3n^2}\Big(n-C_{\alpha}\Big).
\end{equation}
According to Eqs. (\ref{30}) and  (\ref{35}), we obtain  the following expression for the EoS of matter:
\begin{equation}\label{36}
w_m = -1 + \frac{(3-\alpha)(n-C_{\alpha})}{3n(n-C_{\alpha})-\Lambda_0 t^{\displaystyle 3-\alpha}}.
\end{equation}
When $\alpha=1$ and $\Lambda_0=0$, this expression reduces to the well-known one: $\displaystyle w_m=-1+\frac{2}{3n}$.  Putting  the Hubble parameter and the cosmological term (\ref{34}) into Eq. (29), we obtain the total energy density of matter and vacuum as
\begin{equation}\label{37}
\rho(t)+ \Lambda_0=\frac{3n}{t^{\displaystyle 3-\alpha}}(n-C_{\alpha}).
\end{equation}
In the limit $\alpha \to 1$, this implies the standard result: $\rho+ \Lambda_0 = 3n^2/t^2$. Otherwise, when $\alpha \in (0,1)$, the weak energy condition requires $n>C_{\alpha}$, that is simply reduced to $n>0$ in the limit $\alpha \to 1$.

(3) At last we consider a model of the  bouncing universe \cite{Novello}, which can be obtained  from Eqs. (\ref{15}) and (\ref{16}) by putting $\displaystyle \frac{M}{L}=\epsilon\, n,\,\frac{L}{K}=\epsilon \, m$, where $n,m$ are some positive constants, and $\epsilon = \pm \, 1$:
\begin{equation}\label{38}
a(t) = a_0 t^{\displaystyle -\epsilon \, n} \exp \Big(H_0 t^{\displaystyle \epsilon \, m}\Big),\,\,H(t) = \epsilon \, H_0\, m\, t^{\displaystyle \epsilon \,m-1}  -\epsilon\, \frac{n}{t} .
\end{equation}
Note that this solution can also be obtained using the ansatz (\ref{17}), if the coefficients (\ref{14}) obey to the following  relationships: $c_2=\epsilon\,m(1-c_1)-5+2\alpha,\,c_3=nm(1-c_1)$.
Substituting Hubble parameter (\ref{38}) into Eqs. (\ref{27}) and (\ref{28}), we obtain that in this model
\begin{equation}\label{39}
\Lambda_{eff} = \Lambda_0 + \frac{3 C_{\alpha}\epsilon}{t^{\displaystyle 3-\alpha}}\Big(\displaystyle \frac{3-\alpha}{3-\alpha-\epsilon m}mH_0t^{\displaystyle \epsilon m}-n\Big)
\end{equation}
provided that $\alpha+\epsilon m \ne 3$, and the effective EpS is equal to
\begin{equation}\label{40}
w_{eff} = -1  + \frac{\epsilon (3-\alpha-2\epsilon m)}{3(m H_0 t^{\displaystyle \epsilon m}-n)}-\frac{2mn+(1-\alpha)(2-\alpha)}{3(m H_0 t^{\displaystyle \epsilon m}-n)^2}.
\end{equation}
Then, according to Eq. (\ref{30}), the EoS of matter can be represented by the following formula:
\begin{equation}\label{41}
w_m=-1+\frac{\epsilon (3-\alpha-2\epsilon m)(m H_0 t^{\displaystyle \epsilon m}-n)-2mn-(1-\alpha)(2-\alpha)}{3(m H_0 t^{\displaystyle \epsilon m}-n)^2-3\epsilon C_{\alpha}\Big(\displaystyle \frac{3-\alpha}{3-\alpha-\epsilon m}m H_0 t^{\displaystyle \epsilon m}-n\Big)-\Lambda_0 t^{\displaystyle 3-\alpha}}.
\end{equation}
The behavior of the equations of state in accordance with (\ref{40}) and (\ref{41}) for a particular choice of the model parameters is shown in Fig. 3.  The time evolution of the scale factor and the Hubble parameter in Eq. (38) is plotted in Fig. 4  Fig. 5 for $\epsilon = \pm 1$.
\begin{figure}[t]
\includegraphics[width=0.47\textwidth]{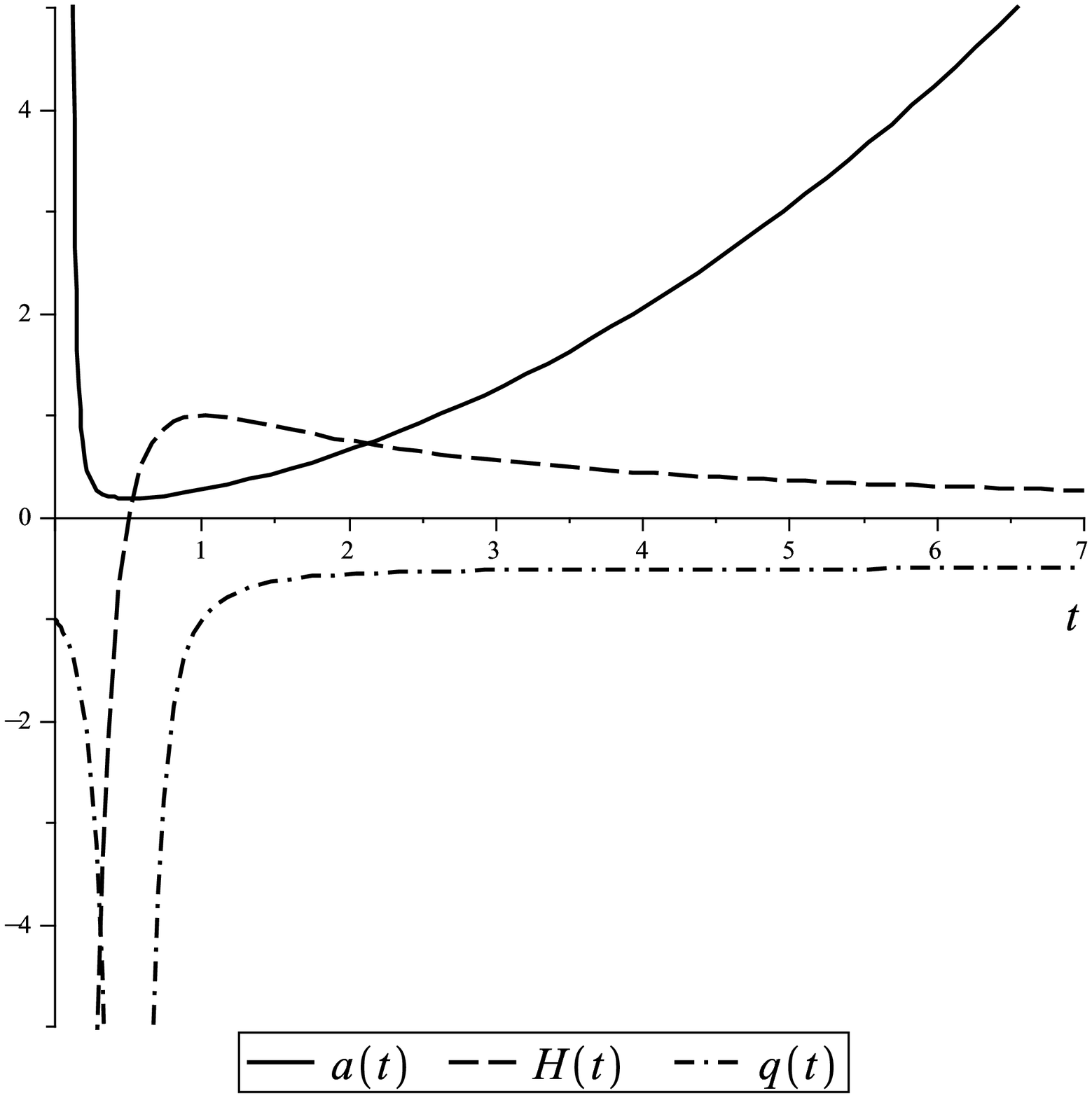} \hfill
\includegraphics[width=0.47\textwidth]{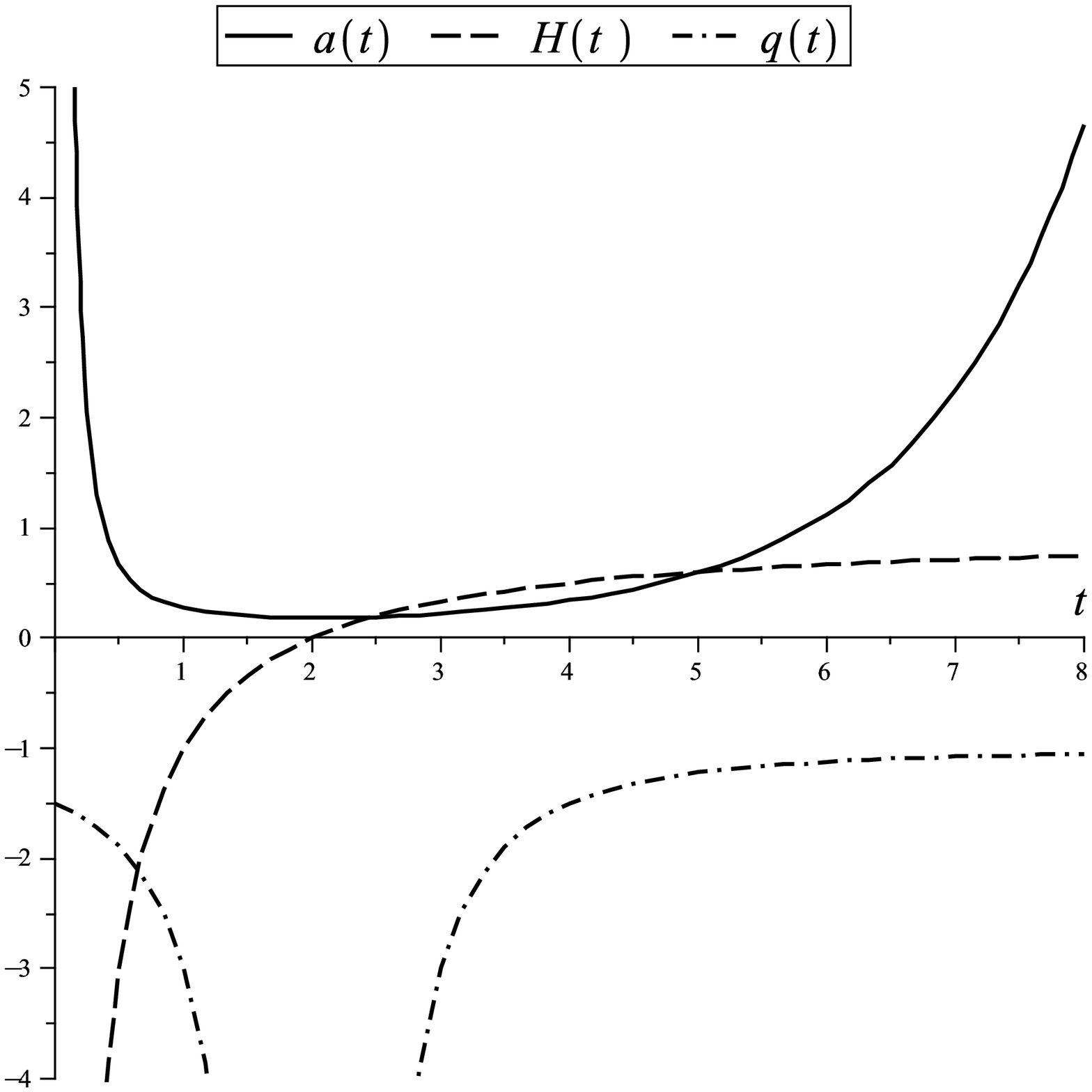}
\parbox[t]{0.47\textwidth}{\caption{Evolution of the bouncing universe model with  $\epsilon=-1$. Here, $n=2,\, m=1,\, H_0=1,\, a_0=0.1$.}\label{fig4}} \hfill
\parbox[t]{0.47\textwidth}{\caption{Evolution of the bouncing universe model with  $\epsilon=+1$. Here, $n=2,\, m=1,\, H_0=1,\, a_0=0.1$.}\label{fig5}}
\end{figure}

\subsection{Models with a constant effective EoS}

Let us note that for a given  effective EoS  $w_{eff}(t)$,  one can obtain the following equation for the Hubble parameter from Eq. (\ref{28}):
\begin{equation}\label{42}
2\dot H+3(1+w_{eff})H^2-\frac{1-\alpha}{t}H+\frac{(1-\alpha)(2-\alpha)}{t^2}=0.
\end{equation}
Finding a generic solution of this equation for an arbitrary function  $w_{eff}(t)$ is not possible. However, one can solve Eq.  (42) for any $w_{eff}=constant$.  The case of $w_m=-1$  (and therefore $w_{eff}=-1$) is discussed  above. However, under this new approach to the effective cosmological term, the situation may change. Indeed, taking into account the solution of Eq. (\ref{20}) represented by the formula (\ref{21}), we have from Eq. (\ref{27}) that
\begin{equation}\label{43}
\Lambda_{eff}=\Lambda_0+\frac{3C_{\alpha}^2}{\displaystyle t^{\displaystyle 3-\alpha}}+\frac{6C_{\alpha}H_0}{t^{\displaystyle (3-\alpha)/2}}.
\end{equation}
Substituting the Hubble parameter (\ref{21}) into Eq. (\ref{29}),  we get rather unforeseen but the well-recognized result for the energy density and pressure of matter: $\rho + \Lambda_0 = 3H_0^2,\,\,p -\Lambda_0= - 3H_0^2$.
We now assume that the constant effective barotropic index  $w_{eff} \ne -1$.
Then it is easy to find a generic solution for Eq. (\ref{42}) in the following form:
\begin{equation}\label{44}
H(t)=\displaystyle \frac{1}{6\,t\,(1+w_{eff})}\Big(3-\alpha + 4 \,n_{\displaystyle (\alpha,w_{eff})}\, \tanh \, \Big[n_{\displaystyle (\alpha,w_{eff})}\,\ln \frac{t}{t_0}\Big]\Big),
\end{equation}
where
\begin{equation}\label{45}
n_{\displaystyle (\alpha,w_{eff})} = \frac{1}{4}\sqrt{(3-\alpha)^2-12(1+w_{eff})(1-\alpha)(2-\alpha)}.
\end{equation}
From Eq.(\ref{44}), one can easily obtain  the scale factor as
\begin{equation}\label{46}
a(t)=a_0\, t^{\displaystyle \frac{3-\alpha}{6(1+w_{eff})}} \Big(\cosh\Big[n_{\displaystyle (\alpha,w_{eff})}\ln \frac{t}{t_0}\Big]\Big)^{\displaystyle \frac{2}{3(1+w_{eff})}}.
\end{equation}
The graphs of the functions (\ref{44}), (\ref{46}),  (\ref{9}) and (\ref{28}) for a particular choice of parameters  $\alpha$ and $w_{eff}$ are shown in Fig. 6. As can be seen from the expression (\ref{45}) with $w_{eff}>(3-\alpha)/12C_{\alpha}$,  $n_{\displaystyle (\alpha,w_{eff})}$ becomes imaginary, and  the expansion (\ref{46}) becomes cyclical.

\begin{figure}[h]
\centering
\includegraphics[width=0.47\textwidth]{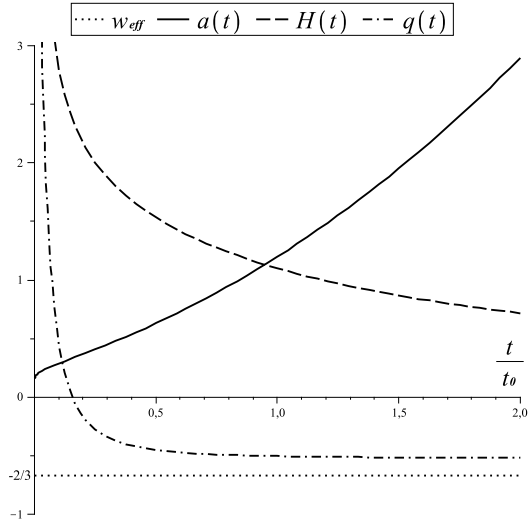}\\
\caption{Evolution of the model $w_{eff}=-2/3$. Here, $\alpha=0.8$ and $a_0=1.2$.}
\label{fig6}
\end{figure}

\subsection{Models from a phenomenologically given cosmological term $\Lambda_{eff}(t)$}

As follows from equation (\ref{27}), a wide class of solution for our model can
be obtained on the basis of the phenomenological laws of evolution of the cosmological term, understanding by this its effective value.  Even the laws which widely discussed in the literature but are not included in the ansatz (\ref{17}) can be considered. Indeed, substituting the effective cosmological term   of the form $\Lambda_{eff}(t) = L(t,a,H)$, where $L(t,a,H)$ is a differentiable function, into Eq. (\ref{27}), we obtain after differentiation with respect to time:
\begin{equation}\label{47}
\frac{\partial L}{\partial t} +\frac{\partial L}{\partial a} \dot a + \frac{\partial L}{\partial H} \dot H +3(1-\alpha)(2-\alpha) \frac{H}{t^{\displaystyle \,3- \alpha}} =0
\end{equation}
This equation can be considered as the main for searching  $a(t)\Rightarrow H(t)$ or $H(t)\Rightarrow a(t)$. After that,  the rest parameters of this model can be obtained from Eqs. (\ref{28})-(\ref{30}).
In some cases depending on the specific function $\Lambda_{eff}(t) = L(t,a,H)$ , the solution for $a(t)$ or $H(t)$ can be found from Eq. (\ref{47}), in some cases - even algebraically.  Let us consider just one illustrative example, making use of the phenomenological law $L=\beta H^2$ which is often discussed in the literature (see e.g.  \cite{Overduin}, \cite{Sahni}).  Then Eq.  (\ref{47}) can be reduced to the following one:
$$
2\beta \dot H +3(1-\alpha)(2-\alpha)t^{\displaystyle \alpha-3}=0.
$$
Integrating the latter and substituting the result into Eq. (\ref{27}) ), together with $\Lambda_{eff}=\beta H^2$, we obtain
\begin{equation}\label{48}
H(t) = \frac{3(1-\alpha)}{2\beta}\cdot \frac{1}{t^{\displaystyle 2-\alpha}} +H_0,
\end{equation}
where $\Lambda_0=\beta H_0^2$. It follows that the scale factor of this model is of the following expression:
\begin{equation}\label{49}
a(t)=a_0 \exp \Big(H_0 t -\frac{3}{2\beta} t^{\displaystyle \alpha-1}\Big),
\end{equation}
where $a_0$ is a constant of integration, and $\alpha \in (0,1)$.
Making use of Eqs. (\ref{48}), (\ref{49}) and $\Lambda_{eff}=\beta H^2$, it is easy to find the deceleration parameter (\ref{9}),
\begin{equation}\label{50}
q(t)=-1+\frac{6\beta (1-\alpha)(2-\alpha)\,t^{\displaystyle 1-\alpha}}{\Big[3(1-\alpha)+2\beta H_0\,t^{\displaystyle 2-\alpha}  \Big]^2},
\end{equation}
the effective EoS (\ref{28}),
\begin{equation}\label{51}
w_{eff}(t)=-1+\frac{2\beta (1-\alpha)(2-\alpha)\,t^{\displaystyle 1-\alpha}}{3\Big[3(1-\alpha)+2\beta H_0\,t^{\displaystyle 2-\alpha}  \Big]}\left(1+\frac{2(2-\alpha)(3-\beta)\,t^{\displaystyle 1-\alpha})}{3(1-\alpha)+2\beta H_0\,t^{\displaystyle 2-\alpha}}\right),
\end{equation}
and the EoS of matter (\ref{30})
\begin{equation}\label{52}
w_m(t)=-1+\frac{3(1+w_{eff})}{3-\beta \,t^{\displaystyle 1-\alpha}},
\end{equation}
where $w_{eff}$ is represented by Eq. (\ref{51}).
\begin{figure}[t]
\includegraphics[width=0.47\textwidth]{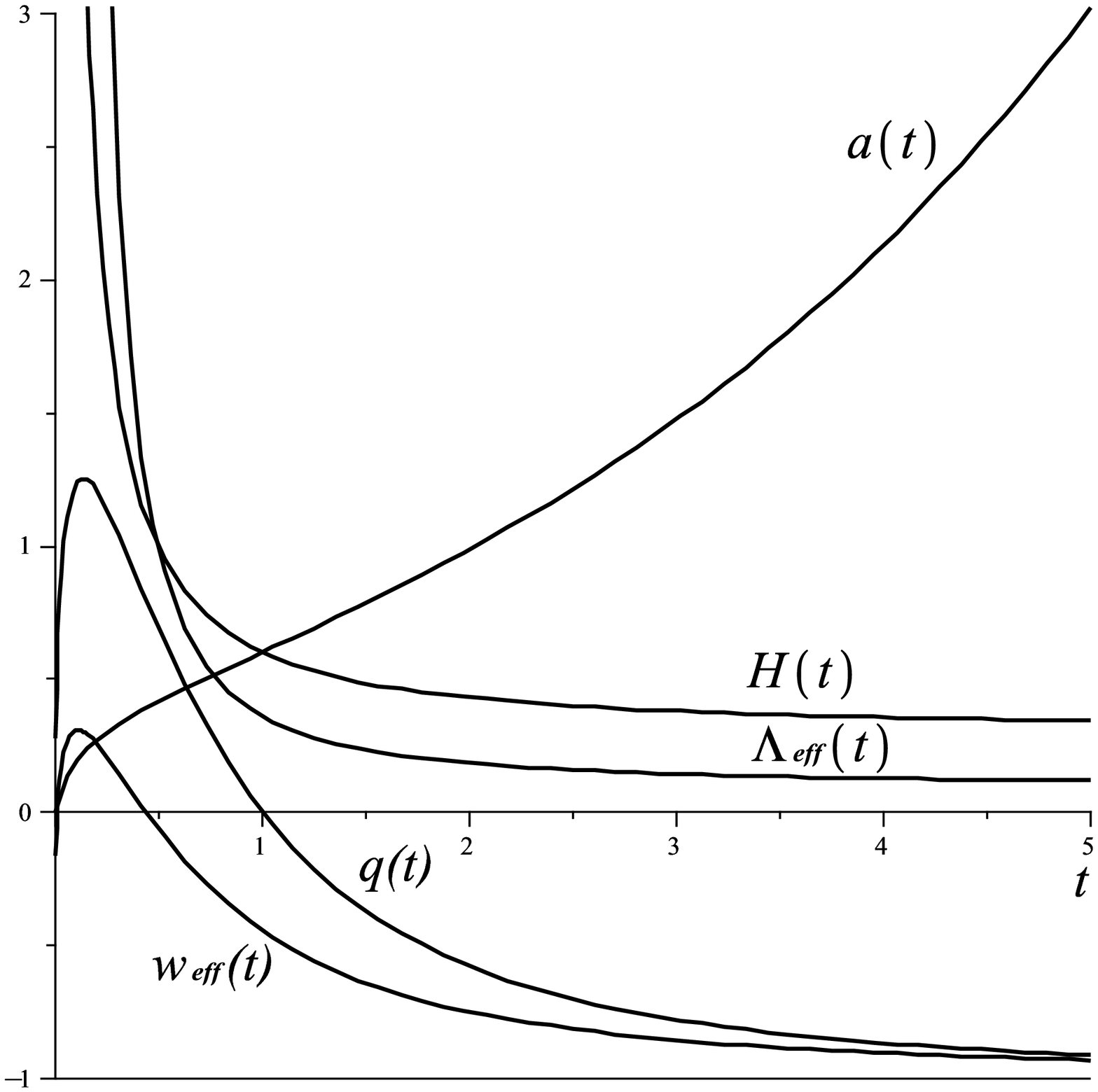} \hfill
\includegraphics[width=0.47\textwidth]{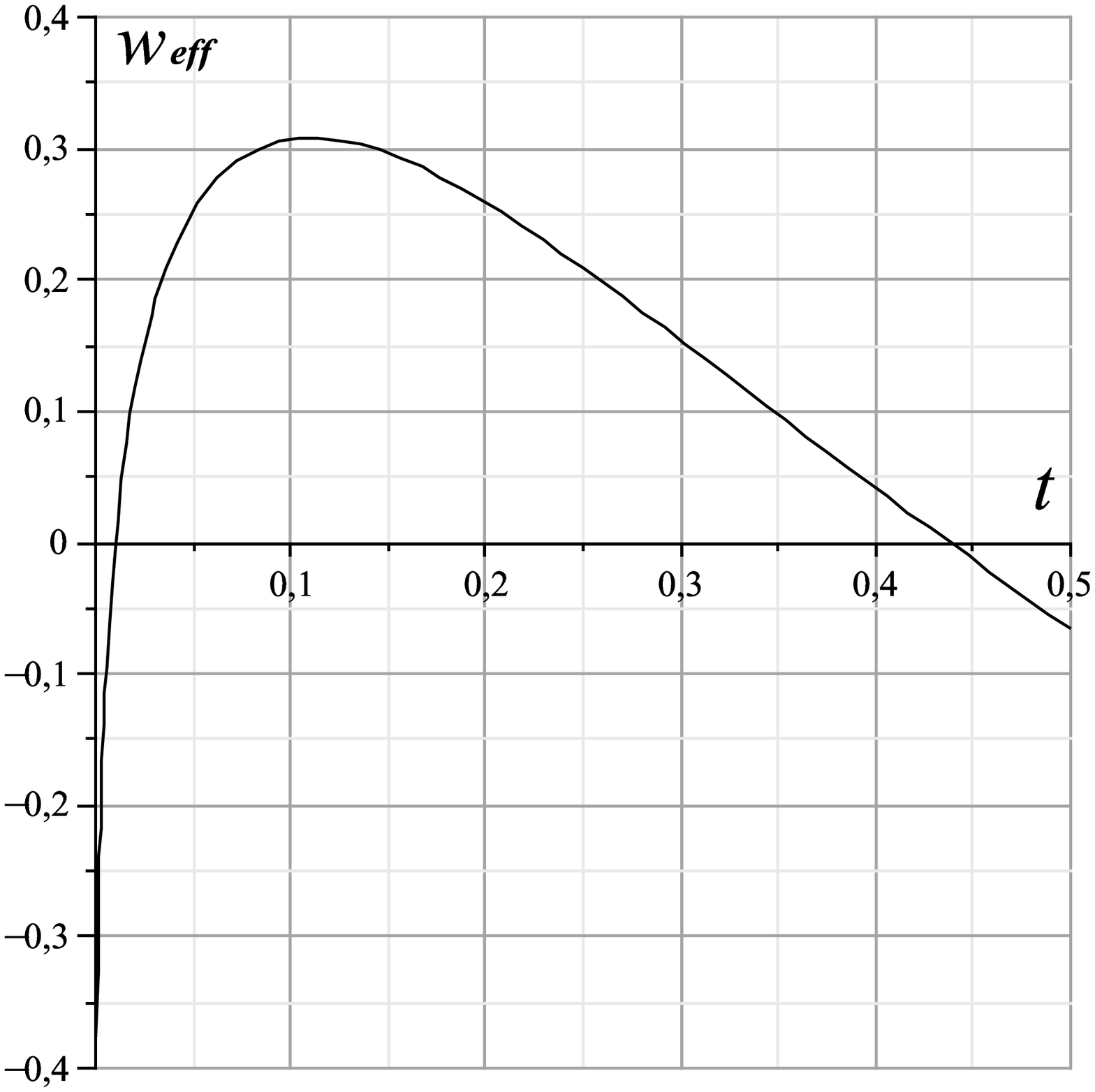}
\parbox[t]{0.47\textwidth}{\caption{Evolution of the model $\Lambda_{eff} = \beta H^2$. Here, $\alpha=0.8$,\, $\beta=1,\,H_0=0.3$ and $a_0=2$.}\label{fig7}} \hfill
\parbox[t]{0.47\textwidth}{\caption{A fragment of Fig. 7 which shows $w_{eff}$ within a shorter interval of time.}\label{fig8}}
\end{figure}
The evolution of this model according to Eqs.  (\ref{48}) - (\ref{51}) for the certain values
of parameters is shown in Fig. 7. It is easy to find that the effective EoS $w_{eff}$
and the deceleration parameter $q$ start at  $w_{eff}(0)=q(0)=-1$ at the initial time
and asymptotically tend to the same value:  $w_{eff}(t \to \infty)=q(t \to \infty)=-1$.
However, during a certain time interval, the effective EoS $0<w_{eff}<1$. Furthermore,   the expansion slows down even for a longer period of time when the deceleration parameter becomes positive.
After that, the expansion again accelerates. It is interesting that the effective EoS $w_{eff}\approx 1/3$ in its maximum . In our view, all these features of the model bring it closer to the realistic scenarios that are widely discussed at present. As one can see, the tuning of this model is possible by means of several parameters, such as $\alpha,\, \beta,\, H_0$ and $a_0$.

\section{Conclusion}

Thus, we have proposed a new cosmological model constructed from the fractional action functional, provided that the fractional order could apply only to the gravitational sector of the effective action. Since in this model only the Einstein-Hilbert action  is subject to modification, the continuity equation, that is the energy-momentum conservation law , has its standard form. Making use of a rather general ansatz for the dynamical cosmological term and some assumptions about the equation of state, we have obtained a lot of exact solutions for the field equations of this model. In any case, the behavior of our model demonstrates its significant difference from the corresponding standard model.  That fact is probably a consequence of the fractal nature of space-time. (see for example \cite{1Calcagni}, \cite{2Calcagni}).

As one can see, our model may posses very interesting  feature:  it is able to evolve cyclicly, that attracts attention of many researchers for the recent years (see \cite{Novello} for a review). In our view, the idea of a kinematically induced cosmological term (by the Hubble parameter), proposed in this paper, is of particular attraction. We have shown in a specific example that the model, based on the difference in weight functions in the action of the gravity and matter, can lead to rather realistic regime of  expansion of the universe.
Obviously, subsequent studies in this area imply the geometric testing of solutions \cite{Starobinsky} and determination of restrictions on the constructed models by the observational data \cite{Riess}, \cite{Perlmutter}.

\end{document}